\documentclass[letterpaper]{article} 
\usepackage{aaai25}  
\usepackage{times}  
\usepackage{helvet}  
\usepackage{courier}  
\usepackage[hyphens]{url}  
\usepackage{graphicx} 
\usepackage{natbib}  
\usepackage{caption} 
\usepackage{algorithm}
\usepackage{algorithmic}
\usepackage{amssymb}

\usepackage{newfloat}
\usepackage{listings}
\DeclareCaptionStyle{ruled}{labelfont=normalfont,labelsep=colon,strut=off} 
\floatstyle{ruled}
\newfloat{listing}{tb}{lst}{}
\floatname{listing}{Listing}
\usepackage{tabularx}
\usepackage{booktabs}
\usepackage{graphicx}

\usepackage{amsmath}
\title{Technique Inference Engine:\\ A Recommender Model to Support Cyber Threat Hunting}
\author{
    Matthew J. Turner,
    Mike Carenzo,
    Jackie Lasky,\\
    James Morris-King,
    James Ross
}
\affiliations{
    MITRE Center for Threat-Informed Defense\\

    7525 Colshire Dr\\
    McLean, VA 22102 USA\\
    \{mjturner, mcarenzo, jlasky, jamesmk, rossj\} @mitre.org
}

\begin{document}

\maketitle

\begin{abstract}
	\emph{Cyber threat hunting} is the practice of proactively searching for latent threats in a network.  Engaging in threat hunting can be difficult due to the volume of network traffic, variety of adversary techniques, and constantly evolving vulnerabilities.  To aid analysts in identifying techniques which may be co-occurring as part of a campaign, we present the Technique Inference Engine, a tool to infer tactics, techniques, and procedures (TTPs) which may be related to existing observations of adversarial behavior.  We compile the largest (to our knowledge) available dataset of cyber threat intelligence (CTI) reports labeled with relevant TTPs.  With the knowledge that techniques are chronically under-reported in CTI, we apply several implicit feedback recommender models to the data in order to predict additional techniques which may be part of a given campaign.  We evaluate the results in the context of the cyber analyst's use case and apply t-SNE to visualize the model embeddings.  We provide our code and a web interface.
\end{abstract}

%
\begin{links}
    \link{Code}{https://github.com/center-for-threat-informed-defense/technique-inference-engine}
    \link{Web Interface}{https://center-for-threat-informed-defense.github.io/technique-inference-engine/#/}

\end{links}

\section{Introduction}\label{sec:introduction}

Cyber attacks have become nearly ubiquitous and threaten economic interests, critical infrastructure, and public health and safety\cite{bitcoin}\cite{colonial_pipeline}\cite{wannacry}.  In the light of such a pernicious threat, governments and organizations have turned to \emph{cyber threat hunting}, or the process of analyzing cyber threat intelligence (CTI) in the context of an organization's risk in order to preemptively identify and stop cybercrime.  Analysts often collect thousands or even millions of data points from network traffic and other digital behavior within an organization.  One cyber threat hunting workflow involves the following steps:

\begin{enumerate}
    \item Identify anomalies in network traffic or other data, called indicators of compromise (IoCs).
    \item Extract the tactics, techniques, and procedures (TTPs) associated with each IoC.  This is often performed by mapping IoCs to the MITRE ATT\&CK\textregistered framework\cite{attack}.
    \item Utilize characteristics of observed adversary behavior to hunt for potentially unobserved TTPs.
\end{enumerate}

We aim to support analysts in this third mission.  Our goal is, given a set of observed ATT\&CK techniques from a campaign, to predict the next most likely techniques in that campaign, thus empowering analysts to hunt for additional adversary behavior.  This problem is made more challenging by the fact that TTPs are chronically underreported in CTI due to analyst bias, potentially undiscovered TTPs in a campaign, and undisclosed proprietary or sensitive information.  For this reason, we apply multiple implicit feedback recommender models that can handle the lack of \emph{negative examples} of TTPs, or training examples in which a particular technique was definitely not part of a campaign.  Our contributions are as follows:

\begin{enumerate}
    \item Compiling the largest (to our knowledge) publicly available dataset of CTI reports tagged with their relevant MITRE ATT\&CK techniques.
    \item Formulating the problem as an implicit feedback recommender task and evaluating multiple models to identify the best performing algorithm for analysts' use case.
    \item Developing an easy-to-use web interface in which analysts can enter observed TTPs and receive a ranked list of techniques the model deems most likely to be associated with the given campaign.
    \item Visualize the learned report embeddings and identify avenues for future research on adversary characterization.
\end{enumerate}

We proceed as follows: first, we discuss the background and related work.  Next, we define our model and experimental setup and discuss and visualize results.  Finally, we draw conclusions and identify potential future research questions.

\section{Related Work}

\subsection{Cyber threat hunting}

Traditional cyber threat hunting tools include Endpoint Detection and Response, or automatic endpoint monitoring for anomalies, and Intrusion Detection Systems, which aim to identify network incursions\cite{cyber_hunting_overview}.  Because these tools solely act as asentinels for adversary behavior, they are not effective against Advanced Persistent Threats (APTs), which can lie dormant on a network months after intrusion.  They also require analysts to group and analyze the disparate IoCs reported across various tools, although Security Information and Event Management software can help\cite{cyber_hunting_overview}.

To move from the simple identification of threats to a proactive response, organizations have turned to predictive analytics and artificial intelligence and machine learning (AI/ML).  \citeauthor{sightings} compiled over 1.6 million IoCs over a period of 2 years and related them to the ATT\&CK framework.  From this, they were able to glean a list of the top 15 observed techniques as well as frequently co-occurring techniques in time.  Although time relation does not guarantee two techniques were part of the same campaign, this work does raise awareness for potential adversary behavior.

Further efforts to apply predictive analytics to cyber hunting have focused on AI/ML applications.  Taking a direct approach to code itself, prior work has explored malware detection\cite{ml_for_malware_detection},  mitigating code obfuscation using LLMs\cite{code_obfuscation}, and automatic vulnerability detection\cite{protocol_fuzzing}.

Another vein of research has attempted to model adversary behavior in order to provide predictive analytics.  \citeauthor{rl_cyber_hunting} and \citeauthor{hemberg-journal} simulated adversarial agents in a cyber game to model the best responses of both adversaries and defenders as Nash equilibria.  Other work has focused on adversary emulation, or the process of characterizing an adversary as a red team and observing its behavior.  CALDERA, introduced by \citeauthor{caldera_paper}, is an open-source adversary emulation platform serviced by \citeauthor{caldera}.  \citeauthor{automatic_adversary_emulation} provides an algorithmic approach to automatic adversary emulation, eliminating the need for expensive penetration testing.  \citeauthor{rl_aep} applied reinforcement learning agents to learn more effective emulation plans.  Although each of these approaches can support threat hunting via observation of simulated adversary behavior, they are primarily attuned to bolstering resilience against future attacks rather than studying the behavior of an already engaged adversary.

\subsection{Recommender systems}

\emph{Collaborative filtering} recommender systems seek to recommend an \emph{item} $i$ to an \emph{entity} (ie user) $u$ based on characteristics gleaned from all observed item-entity interactions, such as views, reviews, etc.  Two common collaborative filtering approaches include matrix factorization (MF) and deep learning models.

MF models aim to embed information about entities and items in a (smaller) latent space which can be used to make predictions based on a variety of similarity metrics, such as dot product or cosine similarity.  Variations for \emph{implicit feedback} are applied when the dataset contains only positive examples (entity $u$ \emph{has interacted} with item $i$) and are usually associated with situations where there is ambiguity as to the reason for the absence of an entity-item interaction.  For example, users may not have purchased a product from an online retailer either because they didn't like the product or because they didn't know it existed.  Implicit feedback recommendation problems may be solved by negative sampling or non-sampling approaches\cite{bayesian_personalized_ranking}\cite{non_sampling}.  A summary of these two methods is presented in Table \ref{tab:negative_vs_non_sampling}.

\begin{table*}[t]
    \centering
    \begin{tabularx}{\textwidth}{lp{0.2\linewidth}Xp{0.2\linewidth}}
         \toprule
         Approach & Objective & Approach & Example \\
         \midrule
         Negative sampling & Maximize log-likelihood of ranking. & Positive examples should rank higher than unobserved negaives among predictions. & Bayesian Personalized Ranking\cite{bayesian_personalized_ranking} \\
         Non-sampling & Minimize training loss. & Weight negatives less than positives in training loss. & Weighted Matrix Factorization\cite{non_sampling} \\
         \bottomrule
    \end{tabularx}
    \caption{A comparison between negative sampling and non-sampling approaches to implicit feedback recommender systems.  Although their performance is often competitive, non-sampling approaches can achieve compeititve results with faster runtimes as they don't need to sample from a distribution\cite{negative_vs_non_sampling}.}
    \label{tab:negative_vs_non_sampling}
\end{table*}

\section{Experimental Setup}\label{sec:experimental_setup}

\subsection{Dataset}

We compile the largest (to our knowledge) available dataset of CTI reports mapped to ATT\&CK techniques..  A summary of the public and private data sources used may be found in Table \ref{tab:data_sources}.  In total, the dataset consists of 43,899 technique observations across 6,236 CTI reports.  We achieved 96\% coverage of ATT\&CK Enterprise 15.0 framework, with 611 of the 637 parent and sub-techniques represented.

\begin{table*}[t]
    \centering
    \begin{tabularx}{\textwidth}{Xll}
         \toprule
         Data Source & Description & \# Reports \\
         \midrule
         Threat Report ATT\&CK Mapper (TRAM) \cite{tram} & Manually annotated dataset of CTI. & 149 \\
         ATT\&CK Flows\cite{attack_flows} & CTI manually annotated with sequences of adversary behavior. & 29 \\
         Adversary Emulation Plans\cite{adversary_emulation_plans} & Library of emulation plans from real-world CTI. & 10 \\
        ATT\&CK Campaigns\cite{attack_campaigns} & Campaigns from the MITRE ATT\&CK framework. & 22 \\
         Participant Shared Data & Contributed data from project participants. & 1 \\
         OpenCTI Data & MITRE internal OpenCTI repository of open-source threat reports. & 6025 \\
         \bottomrule
    \end{tabularx}
    \caption{Summary of data sources present in our compiled dataset.  The majority of the data comes from OpenCTI, a repository compiled by a team of MITRE analysts including a quality review process.}
    \label{tab:data_sources}
\end{table*}

\subsection{Model}

We begin with a set of $m$ CTI reports $E$ and $n$ items $I$.  Our goal is, given observations of campaigns and their associated techniques, to develop a model which can infer additional techniques which may be part of a new campaign.  To that end, we model this problem as a collaborative filtering recommender task.  More formally, given a set of observations of entity-item (or in our case, campaign-technique) interactions $A_{obs} : E \times I$, we seek to, for a new set of technique observations $P : I$, infer a ranking of additional techniques which may belong to that set.

Define $A$ as the (sparse) $mxn$ matrix containing the observations from $A_{obs}$.  In our case, $A$ contains 1's if a technique were observed in a particular campaign, but is mostly sparse.  We note that just because a technique is not tagged to a report does not mean it was not present in the underlying CTI for the reasons stated in the Introduction—this is in essence the implicit feedback problem.  In other words, there is ambiguity as to whether the sparse entries in $A$ result from techniques which were actually not part of the campaign, or whether they were not reported for other reasons.

Our task is to, given the set of observed techniques in a new campaign $P$, rank each technique $i \in I$ in order of relevance to that report.  We compare the results of both negative and non-sampling on this problem.  For negative sampling, we use Bayesian Personalized Ranking (BPR)\cite{bayesian_personalized_ranking}.  For non-sampling, we look to Weighted Matrix Factorization (WMF)\cite{non_sampling}.  We provide our own implementations of both algorithms, as well as a brief summary of each approach below.

\subsubsection{Weighted Matrix Factorization}

The classic matrix factorization approach to collaborative filtering seeks to decompose $A$ into latent factors $U$ and $V$:

\begin{equation}\label{eq:matrix_factorization}
    A = UV^T
\end{equation}

where $A \in \mathbb{R}^{mxn}$ and $U, V \in \mathbb{R}^{mxd}$, where $d$ is the embedding dimension $d < min(\{m, n\})$.  As a non-sampling method, WMF seeks to minimize the training loss over all samples simultaneously, while weighting positive, observed examples of techniques in a report more than examples where a technique was not observed in a report.  In particular, we have the following objective function:

\begin{multline*}
    \min_{U, V} \sum_{(i, j) \in A_{obs}} \left( A_{ij} - \langle U_i, V_j \rangle \right)^2 + w \sum_{(i, j) \notin A_{obs}} \langle U_i, V_j \rangle \\ + \lambda \left( \sum_i ||U_i||^2 + \sum_j ||V_j||^2 \right)
\end{multline*}

where $\lambda$ is some regularization parameter tuned by cross validation.  We use the formulation from \citeauthor{google_recommender_systems}, but note this is equivalent to setting the unobserved entries in $A$ to some weight $1-w$, where $0 \leq w < 1$.  In essence, we fill in the matrix $A$ with 0's for the unobserved entries, but associate them with a lower confidence in the objective.  Although this minimization problem is not convex, fixing one set of latent factors (ie $U$) results in a quadratic problem in $V$, and alternating between the two is guaranteed to converge to the optimum\cite{non_sampling}.

\subsubsection{Bayesian Probabilistic Ranking}

On the other hand, BPR, a negative sampling method, proposes to perform a Bayesian analysis to maximize the likelihood of a correct ranking based on the assumption that observed entity-item interactions should always be ranked higher than unobserved ones.  To do this, the authors define a dataset $D$ such that the elements of $D$ are triples of the form $(i, j, k)$ where $(i,j$ is an observed entity-item interaction and item $k$ did not interact with $i$.  More formally, the authors define a dataset $D : E \times I \times I$ s.t. $D := \left\{(i, j, k) | (i, j) \in A_{obs} \wedge (i, k) \notin A_{obs}\right\}$.  Sampling uniformly from this distribution, we maximize

\begin{displaymath}
\begin{split}
BPR\_OBJ & = ln p(\Theta | \sigma(x(i, j) - x(i, k))) \\
& =ln p(\sigma(x(i, j) - x(i, k) | \Theta) \\
& =ln \prod_{(i, j, k) \in D} \sigma(x(i, j) - x(i, k)p(\Theta) \\
& =\sum_{(u,i,j) \in D} ln \sigma(x(i, j) - x(i, k)) - \lambda ||\Theta||^2 \\
\end{split}
\end{displaymath}

where $x(u, i)$ represents the prediction of the recommender model for entity u and item i, the sigmoid is added to preserve a total ordering, and $\lambda$ is a regularization parameter tuned via cross validation.  Although $x(u, i)$ could be any (differentiable) prediction model, we use the same matrix factorization model in Equation \ref{eq:matrix_factorization}; therefore, $x(u, i) = \langle U_i, V_j \rangle$.

\subsection{Experiments}\label{sec:experiments}

Consistent with prior work, we begin by selecting 20\% of the data as a test set and leave out an additional 10\% of the training data as a validation set.  We perform hyperparameter tuning via grid search, consistent with prior research\cite{negative_vs_non_sampling}.  We select $recall@K$ and $NDCG@K$ as our primary evaluation metrics.  Define $R_i = [r_i^1, r_i^2, \dots r_i^K]$ as ranked list (by prediction score) of techniques inferred by the model in descending order.  Let $T_i$ be the withheld test set for entity/campaign $i$.  Then we have:

\begin{displaymath}
    recall@K = \frac{1}{|E|} \sum_{i=1}^m \frac{\sum_{k=1}^K [[r_i^k \in T_i]]}{|T_i|}
\end{displaymath}

where $[[x]]$ is the indicator function returning 1 if $x$ is True, 0 otherwise.  We also define $DCG@K$:

\begin{displaymath}
    DCG@K = \frac{1}{|E|} \sum_{i=1}^m \sum_{k=1}^K \frac{[[r_i^k \in T_i]]}{log_2 (k+1)}
\end{displaymath}

$NDCG@K$ is simply $DCG@K$ normalized by the maximum value obtained from the best possible ranking.  

We chose $recall@K$ since it relates to the fraction of relevant techniques our users can expect to see after viewing the first $K$ predictions.  We attend to $NDCG@K$ as it takes into account the relative rankings of those relevant predictions.  We select the hyperparmaters with the highest $recall@20$ on the validation data, since our web interface will by default display 20 results.  Our hyperparameter combinations for BPR and WMF may be found in Tables \ref{tab:bpr_hyperparameters} and \ref{tab:wmf_hyperparameters}, respectively.  We measure both dot product and cosine similarity for all hyperparameter combinations.  Dot product similarity favors popular items in predictions, while cosine similarity normalizes this impact to some degree\cite{google_recommender_systems}.

\begin{table*}[t]
    \centering
    \begin{tabular}{l|l|c}
        \toprule
        Hyperparameter & Possible Values & Best Value \\
        \midrule
        Embedding dimension & 4, 8, 16, 32 & 16 \\
        Learning rate & 0.00001, 0.00005, 0.0001, 0.001, 0.005, 0.01, 0.02, 0.05 & 0.02\\
        Regularization coefficient & 0.0, 0.0001, 0.001, 0.01 & 0.01 \\
        \bottomrule
    \end{tabular}
    \caption{Hyperparameter combinations for BPR.  We adopt the learning rate options from \cite{negative_vs_non_sampling} while adding the values below 0.001.  The regularization coefficient options are also from \cite{negative_vs_non_sampling} with the addition of 0.00001.  Unlike \cite{negative_vs_non_sampling}, we additionally test multiple embedding dimensions.}
    \label{tab:bpr_hyperparameters}
\end{table*}

\begin{table*}[t]
    \centering
    \begin{tabular}{l|l|c}
        \toprule
        Hyperparameter & Possible Values & Best Value \\
        \midrule
        Embedding dimension & 4, 8, 16, 32 & 4 \\
        $c$ & 0.0001, 0.001, 0.005, 0.01, 0.05, 0.1, 0.3, 0.5, 0.7 & 0.001 \\
        Regularization coefficient & 0.0, 0.00001, 0.0001, 0.001, 0.01 & 0.00001 \\
        \bottomrule
    \end{tabular}
    \caption{Hyperparameter combinations for WMF.  $c$ represents the weight for negative samples as compared to positive samples which take weight 1.  We take the values for $c$ from \cite{negative_vs_non_sampling} with the addition of 0.0001.  The regularization coefficient options are also from \cite{negative_vs_non_sampling} with the addition of 0.00001.}
    \label{tab:wmf_hyperparameters}
\end{table*}

\section{Results}\label{sec:results}

In line with prior research\cite{negative_vs_non_sampling}, we run each model 5 times and report the average $recall@K$ and $NDCG@K$ in Table \ref{tab:results}.  As a baseline, we also report results for a model TopTechniquesRecommender, which recommends techniques in ranked order of their frequency in the dataset.  

\begin{table*}[t]
    \centering
    \begin{tabular}{l|c|c|c|c|c|c}
        \toprule
         Model & Recall@10 & Recall@20 & Recall@50 & NDCG@10 & NDCG@20 & NDCG@50 \\
         \midrule
         TopTechniquesRecommender & 0.2285 & 0.3419 & 0.5646 & 0.1563 & 0.1991 & 0.2672 \\
         \midrule
         WMF & \textbf{0.2613} & \textbf{0.4037} & 0.6258 & \textbf{0.1697} & \textbf{0.2232} & \textbf{0.2951} \\
         \midrule
         BPR & 0.2382 & 0.4001 & \textbf{0.6579} & 0.1434 & 0.2018 & 0.2815 \\
         \bottomrule
    \end{tabular}
    \caption{Results for BPR, WMF, and the TopTEchniquesRecommender.  WMF outperforms in every instance.}
    \label{tab:results}
\end{table*}

We note that WMF and BPR achieve similar results, with WMF outperforming in $recall@10$, $recall@10$ and all $NDCG$ metrics.  However, BPR slightly overperformed in $recall@50$.  This is likely because, as mentioned in Table \ref{tab:negative_vs_non_sampling}, BPR optimizes to rank observed examples over unobserved in the loss function.  This may bias its predictions toward popular techniques that are observed frequently.  We note that, similarly, the TopTechniquesRecommender performs at the closest recall to WMF, 90.2\%, at $k=50$, further solidifying the idea that predicting high-frequency techniques has greater benefit at larger values of $K$.  A comprehensive study of negative and non-sampling methods in implicit feedback recommender systems\cite{negative_vs_non_sampling} found that WMF generally excels against BPR.  This becomes even more significant when we note that WMF is able to obtain very similar results with a much smaller embedding dimension of 4 versus 16, leading to less model complexity and faster runtimes.  In our implementations, WMF trains considerably faster (on order of minutes versus hours), although we do run BPR with a batch size of 1, as opposed to 512 as used in previous research\cite{negative_vs_non_sampling}.  WMF's faster performance due both to the speed of the algorithm and the need to run it for fewer iterations to achieve convergence.  In our experiments 25 epochs was sufficient for WMF training, while BPR required 100 epochs to converge.  The combination of these factors motivates us to apply WMF as the backbone for our web implementation.  In fact, we are able to perform the WMF computations locally on the user's browser, eliminating the need for a backend server at all.

Both models outperform the TopTechniquesRecommender, with WMF excelling by approximately 18\% for $recall@20$ and 12\% for $NDCG@20$.  This better relative performance for recall is expected due to the heavy presence of popular techniques in the dataset.  For example, the most occurring technique, T1059, is present in 2,069 of the 6,236 CTI.  The TopTechniquesRecommender can score highly on NDCG simply by making sure that technique, along with other popular techniques, are ranked highly.  Despite this fact, the modest outperformance of WMF suggests that the system is learning characterizations of reports and techniques that allow it to outperform frequency alone.  Finally, we note that dot product similarity outperformed cosine in all instances, likely due to the same impact of popular techniques.

\begin{figure*}
    \centering
    \includegraphics[width=0.7\linewidth]{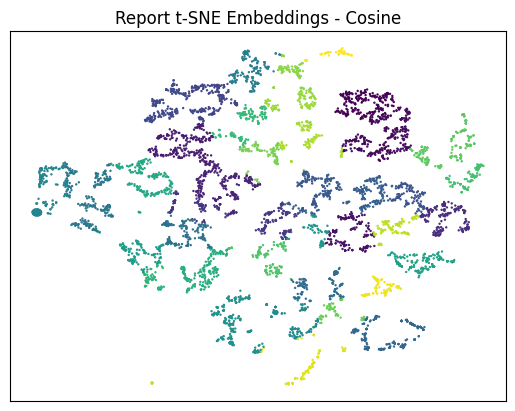}
    \caption{t-SNE visualization of WMF report embeddings using the cosine distance metric and a perplexity of 30.  Colors are based on clusters of the data using MeanShift with a bandwidth of 10\cite{mean_shift}.}
    \label{fig:t-sne}
\end{figure*}

We can gain insight into the manner in which the model characterizes the reports by applying t-SNE, a technique to visualize points via dimensionality reduction, to the report embeddings\cite{t-sne}.  We apply t-SNE on the report embeddings for WMF with perplexity 30 and a cosine distance metric, which reduces the impact of popular techniques.  The results are shown in Figure \ref{fig:t-sne}.  This analysis may be useful both for identifying similar reports to those entered by analysts and for learning from characteristics of particular adversaries.

\section{Web Interface}\label{sec:interface}

\begin{figure*}
    \centering
    \includegraphics[width=0.6\linewidth]{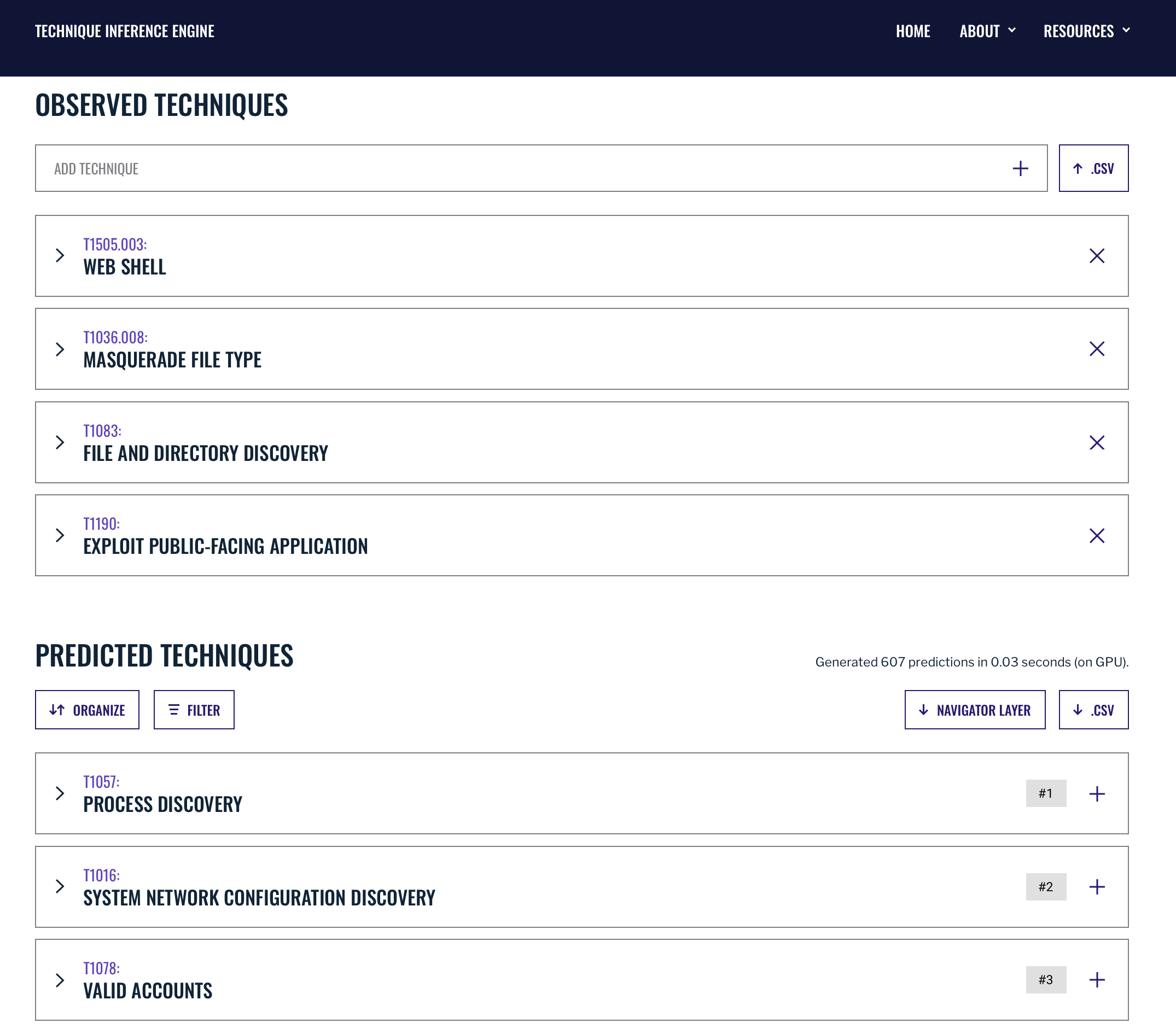}
    \caption{Technique Inference Engine web interface.  Users may add observed techniques using the "ADD TECHNIQUE" box and retrieve a list of predicted technqiues which may be sorted, filtered, and exported to an ATT\&CK Navigator layer or .csv file\cite{attack_navigator}.  In this example, the Technique Inference Engine is able to predict 6 additional techniques present in the MITRE NERVE breach from an input of four techniques\cite{nerve}.  Note that not all predicted techniques are shown.}
    \label{fig:tie-interface}
\end{figure*}

We provide a mature web interface allowing users to input observed techniques and retrieve a list of the top $K=20$ inferred techniques from the recommender model.  We choose $K=20$ as it experimentally provides a decent tradeoff between maintaining a small and relevant prediction set.  A screenshot of the interface may be found in Figure \ref{fig:tie-interface}, and is available at \url{https://center-for-threat-informed-defense.github.io/technique-inference-engine/#/}.

\section{Conclusion}\label{sec:conclusion}

Cyber threat hunting is a complex task made more difficult by the number of adversaries and the breadth of techniques they employ.  The Technique Inference Engine is a tool that can aid analysts in threat hunting by predicting additional techniques that may be part of an adversary's campaign based on observed techniques and previous CTI reports.  

We modeled this problem as an implicit feedback collaborative filtering task and applied both the Bayesian Personalized Ranking and Weighted Matrix Factorization algorithms.  Consistent with prior research, we found a slight edge for WMF in both training time and performance, as measured by $recall@K$ and $NDCG@K$\cite{negative_vs_non_sampling}.

With a $recall@20$ of over 40\%, we believe the model has the accuracy necessary to aid cyber analysts.  If an analyst can enter observed techniques and expect to see 40\% of the additional techniques leveraged by that adversary in their campaign in the first 20 predictions, this is a great step toward hunting additional adversary behavior.  We implement the Technique Inference Engine as an easy-to-use web tool at \url{http://ctid.io/technique-inference-engine}.  We look forward to seeing how threat hunters apply TIE, in concert with other tools, to support their cyber workflows.

\section{Future Work}\label{sec:future_work}

Future work may focus on improving model training or analysis of the results.  To improve model training, existing tools could be applied to augment the dataset via automatic tagging of CTI reports with ATT\&CK techniques\cite{ttp_drill}\cite{tram}.  Some work would be required to adapt these tools to our dataset format and scope, develop a web scraper to gather reports, and verify the effectiveness of this approach.  Any mislabeling could influence the accuracy of the results and the fidelity to real world data.  For the model, an exploration could be done into applying a neural net to create either the report or technique embeddings, at the risk of overfitting.  A more detailed analysis of the results could include gaining insights from the embeddings into similarities between both reports and techniques.  This could potentially open the door to a deeper understanding of adversary behavior.  Finally, work such as \cite{confidence_aware_matrix_factorization} could be extended to provide confidence intervals for implciit feedback recommender systems, allowing us to show a confidence metric for each prediction to users of the tool.

\section{Acknowledgements}
Many thanks to the Center for Threat-Informed Defense research participants who contributed to this project.  Thanks to Mary Parmelee for her review of this work.

\bibliography{references}

\begin{thebibliography}{29}
\providecommand{\natexlab}[1]{#1}

\bibitem[{Applebaum et~al.(2016)Applebaum, Miller, Strom, Korban, and Wolf}]{caldera_paper}
Applebaum, A.; Miller, D.; Strom, B.; Korban, C.; and Wolf, R. 2016.
\newblock Intelligent, automated red team emulation.
\newblock In \emph{Proceedings of the 32nd Annual Conference on Computer Security Applications}, ACSAC '16, 363–373. New York, NY, USA: Association for Computing Machinery.
\newblock ISBN 9781450347716.

\bibitem[{Beerman et~al.(2023)Beerman, Berent, Falter, and Bhunia}]{colonial_pipeline}
Beerman, J.; Berent, D.; Falter, Z.; and Bhunia, S. 2023.
\newblock A Review of Colonial Pipeline Ransomware Attack.
\newblock In \emph{2023 IEEE/ACM 23rd International Symposium on Cluster, Cloud and Internet Computing Workshops (CCGridW)}, 8--15.

\bibitem[{Bhattacharya et~al.(2020)Bhattacharya, Ramachandran, Banik, Dowling, and Bopardikar}]{rl_aep}
Bhattacharya, A.; Ramachandran, T.; Banik, S.; Dowling, C.~P.; and Bopardikar, S.~D. 2020.
\newblock Automated Adversary Emulation for Cyber-Physical Systems via Reinforcement Learning.
\newblock In \emph{2020 IEEE International Conference on Intelligence and Security Informatics (ISI)}, 1--6.

\bibitem[{Chen et~al.(2023)Chen, Ma, Zhang, Wang, Liu, and Ma}]{negative_vs_non_sampling}
Chen, C.; Ma, W.; Zhang, M.; Wang, C.; Liu, Y.; and Ma, S. 2023.
\newblock Revisiting Negative Sampling vs. Non-sampling in Implicit Recommendation.
\newblock \emph{ACM Trans. Inf. Syst.}, 41(1).

\bibitem[{Comaniciu and Meer(2002)}]{mean_shift}
Comaniciu, D.; and Meer, P. 2002.
\newblock Mean shift: a robust approach toward feature space analysis.
\newblock \emph{IEEE Transactions on Pattern Analysis and Machine Intelligence}, 24(5): 603--619.

\bibitem[{CTID(2020)}]{adversary_emulation_plans}
CTID. 2020.
\newblock Adversary {Emulation} {Library}.
\newblock \url{https://mitre-engenuity.org/cybersecurity/center-for-threat-informed-defense/adversary-emulation-library/}.
\newblock Accessed: 2024-07-30.

\bibitem[{CTID(2021)}]{tram}
CTID. 2021.
\newblock Threat {Report} {ATT}\&{CK} {Mapper} ({TRAM}) v1.
\newblock \url{https://mitre-engenuity.org/cybersecurity/center-for-threat-informed-defense/our-work/threat-report-attck-mapper-tram-v1/}.
\newblock Accessed: 2024-07-30.

\bibitem[{CTID(2022)}]{attack_flows}
CTID. 2022.
\newblock Example {Flows} — {Attack} {Flow} v2.2.7 documentation.
\newblock \url{https://center-for-threat-informed-defense.github.io/attack-flow/example_flows/}.
\newblock Accessed: 2024-07-30.

\bibitem[{CTID(2024)}]{sightings}
CTID. 2024.
\newblock Sightings {Ecosystem} v2.0.0 — {Sightings} {Ecosystem} v2.0.0 documentation.
\newblock \url{https://center-for-threat-informed-defense.github.io/sightings_ecosystem/}.
\newblock Accessed: 2024-07-19.

\bibitem[{Fang et~al.(2024)Fang, Miao, Srivastav, Liu, Zhang, Fang, Asmita, Tsang, Nazari, Wang, and Homayoun}]{code_obfuscation}
Fang, C.; Miao, N.; Srivastav, S.; Liu, J.; Zhang, R.; Fang, R.; Asmita; Tsang, R.; Nazari, N.; Wang, H.; and Homayoun, H. 2024.
\newblock Large Language Models for Code Analysis: Do LLMs Really Do Their Job?
\newblock arXiv:2310.12357.

\bibitem[{for Threat Informed~Defense(2024)}]{nerve}
for Threat Informed~Defense, C. 2024.
\newblock {MITRE} {NERVE}.

\bibitem[{Ghafur et~al.(2019)Ghafur, Kristensen, Honeyford, Martin, Darzi, and Aylin}]{wannacry}
Ghafur, S.; Kristensen, S.; Honeyford, K.; Martin, G.; Darzi, A.; and Aylin, P. 2019.
\newblock A retrospective impact analysis of the {WannaCry} cyberattack on the {NHS}.
\newblock \emph{NPJ Digital Medicine}, 2: 98.

\bibitem[{Google(2025)}]{google_recommender_systems}
Google. 2025.
\newblock Matrix factorization {\textbar} {Machine} {Learning}.

\bibitem[{Grobys et~al.(2022)Grobys, Dufitinema, Sapkota, and Kolari}]{bitcoin}
Grobys, K.; Dufitinema, J.; Sapkota, N.; and Kolari, J.~W. 2022.
\newblock What’s the expected loss when Bitcoin is under cyberattack? A fractal process analysis.
\newblock \emph{Journal of International Financial Markets, Institutions and Money}, 77: 101534.

\bibitem[{Hemberg et~al.(2024)Hemberg, Turner, Rutar, and O’reilly}]{hemberg-journal}
Hemberg, E.; Turner, M.~J.; Rutar, N.; and O’reilly, U.-M. 2024.
\newblock Enhancements to Threat, Vulnerability, and Mitigation Knowledge for Cyber Analytics, Hunting, and Simulations.
\newblock \emph{Digital Threats}, 5(1).

\bibitem[{Hu, Koren, and Volinsky(2008)}]{non_sampling}
Hu, Y.; Koren, Y.; and Volinsky, C. 2008.
\newblock Collaborative Filtering for Implicit Feedback Datasets.
\newblock In \emph{2008 Eighth IEEE International Conference on Data Mining}, 263--272.

\bibitem[{Husari et~al.(2017)Husari, Al-Shaer, Ahmed, Chu, and Niu}]{ttp_drill}
Husari, G.; Al-Shaer, E.; Ahmed, M.; Chu, B.; and Niu, X. 2017.
\newblock TTPDrill: Automatic and Accurate Extraction of Threat Actions from Unstructured Text of CTI Sources.
\newblock In \emph{Proceedings of the 33rd Annual Computer Security Applications Conference}, ACSAC '17, 103–115. New York, NY, USA: Association for Computing Machinery.
\newblock ISBN 9781450353458.

\bibitem[{Kulkarni, Ashit, and Chetan(2023)}]{cyber_hunting_overview}
Kulkarni, M.~S.; Ashit, D.~H.; and Chetan, C.~N. 2023.
\newblock A Proactive Approach to Advanced Cyber Threat Hunting.
\newblock In \emph{2023 7th International Conference on Computation System and Information Technology for Sustainable Solutions (CSITSS)}, 1--6.

\bibitem[{Meng et~al.(2024)Meng, Mirchev, Böhme, and Roychoudhury}]{protocol_fuzzing}
Meng, R.; Mirchev, M.; Böhme, M.; and Roychoudhury, A. 2024.
\newblock Large Language Model guided Protocol Fuzzing.
\newblock In \emph{Network and Distributed System Security Symposium}.

\bibitem[{MITRE(2024{\natexlab{a}})}]{caldera}
MITRE. 2024{\natexlab{a}}.
\newblock Caldera.
\newblock \url{https://caldera.mitre.org/}.
\newblock Accessed: 2024-07-27.

\bibitem[{MITRE(2024{\natexlab{b}})}]{attack_campaigns}
MITRE. 2024{\natexlab{b}}.
\newblock Campaigns {\textbar} {MITRE} {ATT}\&{CK}®.
\newblock \url{https://attack.mitre.org/campaigns/}.
\newblock Accessed: 2024-07-30.

\bibitem[{MITRE(2024{\natexlab{c}})}]{attack}
MITRE. 2024{\natexlab{c}}.
\newblock {MITRE} {ATT}\&{CK}®.
\newblock \url{https://attack.mitre.org/}.
\newblock Accessed: 2024-07-19.

\bibitem[{MITRE(2025)}]{attack_navigator}
MITRE. 2025.
\newblock {MITRE} {ATT}\&{CK} {Navigator}.

\bibitem[{Rendle et~al.(2009)Rendle, Freudenthaler, Gantner, and Schmidt-Thieme}]{bayesian_personalized_ranking}
Rendle, S.; Freudenthaler, C.; Gantner, Z.; and Schmidt-Thieme, L. 2009.
\newblock BPR: Bayesian personalized ranking from implicit feedback.
\newblock In \emph{Proceedings of the Twenty-Fifth Conference on Uncertainty in Artificial Intelligence}, UAI '09, 452–461. Arlington, Virginia, USA: AUAI Press.
\newblock ISBN 9780974903958.

\bibitem[{Shaukat, Luo, and Varadharajan(2023)}]{ml_for_malware_detection}
Shaukat, K.; Luo, S.; and Varadharajan, V. 2023.
\newblock A novel deep learning-based approach for malware detection.
\newblock \emph{Engineering Applications of Artificial Intelligence}, 122: 106030.

\bibitem[{Turner, Hemberg, and O'Reilly(2022)}]{rl_cyber_hunting}
Turner, M.~J.; Hemberg, E.; and O'Reilly, U.-M. 2022.
\newblock Analyzing multi-agent reinforcement learning and coevolution in cybersecurity.
\newblock In \emph{Proceedings of the Genetic and Evolutionary Computation Conference}, GECCO '22, 1290–1298. New York, NY, USA: Association for Computing Machinery.
\newblock ISBN 9781450392372.

\bibitem[{van~der Maaten and Hinton(2008)}]{t-sne}
van~der Maaten, L.; and Hinton, G. 2008.
\newblock Visualizing Data using t-SNE.
\newblock \emph{Journal of Machine Learning Research}, 9(86): 2579--2605.

\bibitem[{Wang et~al.(2018)Wang, Liu, Wu, Chen, Liu, Huang, and Huang}]{confidence_aware_matrix_factorization}
Wang, C.; Liu, Q.; Wu, R.; Chen, E.; Liu, C.; Huang, X.; and Huang, Z. 2018.
\newblock Confidence-Aware Matrix Factorization for Recommender Systems.
\newblock \emph{Proceedings of the AAAI Conference on Artificial Intelligence}, 32(1).

\bibitem[{Yoo et~al.(2020)Yoo, Park, Lee, Ahn, Kim, Seo, and Kim}]{automatic_adversary_emulation}
Yoo, J.~D.; Park, E.; Lee, G.; Ahn, M.~K.; Kim, D.; Seo, S.; and Kim, H.~K. 2020.
\newblock Cyber Attack and Defense Emulation Agents.
\newblock \emph{Applied Sciences}, 10(6).

\end{thebibliography}

\end{document}